\documentclass[aps,pra,preprint,groupedaddress]{revtex4}

\usepackage{graphicx,wrapfig}
\usepackage{amsmath}
\usepackage{amsfonts}
\usepackage{amssymb}
\usepackage{epsfig}
\usepackage{subfigure}

\begin{document}


\title{The effect of phase mismatch on second harmonic generation in negative index materials}



\author{ Zh. Kudyshev$^1$, I.~Gabitov$^{2,3}$ and  A.~Maimistov$^4$ }
\affiliation{$^{1}$ Department of Physics, Al-Farabi Kazakh National University, al-Farabi ave., 71, Almaty, 050038,  KAZAKHSTAN\\
$^{2}$Department  of Mathematics, the University of Arizona, 617 N. Santa Rita, Tucson,  AZ 85721-0089, USA\\
$^{3}$Department  of Mathematics, SMU,  3200 Dyer Street, Dallas TX 75275-0156 , USA\\
$^{4}$Department of Solid State Physics and Nanosystems, National
research nuclear university MEPhI, Kashirskoe sh. 31, Moscow,
115409, RUSSIA}


\date{\today}

\begin{abstract}
Second harmonic generation in negative index metamaterials  is
considered.  Theoretical analysis of the corresponding  model
demonstrated significant difference of this phenomenon in
conventional and negative index materials. In contrast to
conceptional materials there is nonzero critical phase mismatch.
The behavior of interacting weaves is dramatically different when phase
mismatch is smaller or greater than critical value.
\end{abstract}


\maketitle

\section{Introduction}
Experimental demonstration of the phenomenon of negative index of
refraction first in the microwave~\cite{SSS:01} and latter in the optical
regime~\cite{SCCYSDK:05,ZFPMOB:05} has stimulated growing interest in
nonlinear properties of negative index materials~\cite{SarShal:07}. This
interest is motivated by specifics on the interaction of
electromagnetic waves with negative index materials. In combination
with a nonlinear response of the optical material to electromagnetic
radiation, this iteration leads to a new nonlinear optical phenomena.
Study of these phenomena is of considerable importance both for
better understanding of fundamentals of electrodynamics of negative
index materials and their applications.  One of the most fundamental
property of negative index material is an opposite directionality of
the Poynting vector, characterizing the energy flux, to the wave vector
$\vec{k}$. On the other hand, the negative index property can be realized
only on particukar wavelength intervals. These two features are
offering a very unusual type of multi-wave interactions, if frequencies
of interacting waves correspond to frequency intervals where optical
material has different signs of refractive index. Multi-wave
interaction must satisfy a phase matching condition, which is
possible only when all wave vectors  are pointed in the same
direction~\cite{Shen:84}. Therefore energy fluxes of the waves with
frequencies corresponding to a negative sign of refractive index
will propagate in opposite direction to those  with
frequencies corresponding to a positive sign of refraction index.

Such effect was suggested for the first time in~\cite{ASBZ:04}, which
considered the particular case of three
waves interaction - second harmonic generation. A solution of the
equation describing   second harmonic generation in the case of exact
phase matching was given in~\cite{PShV:06}. The feasibility of
parametric amplification using three-wave interaction for
compensation  losses in negative index materials was studied
in~\cite{PSh:06}. The dynamics of interacting  wave packets propagating
in negative index materials in the case of second harmonic generation
was considered in~\cite{MGK:07}. It was shown that in contrast to a
weak intensity of pump field, at high intensities a second
harmonic pulse can be trapped by the pump pulse and forced to propagate
in the same direction.

In this paper we investigate second harmonic generation in the presence
of phase-mismatch $\Delta$. This is an important case since
phase-mismatch is more relevant to realistic experimental
conditions. Additionally, it introduces two types of spatial distribution of
second harmonic field intensity along the sample: monotonic and
periodic 1 on the coordinate. Both cases are considered in this
paper. We also studied second harmonic generation near a critical
phase-mismatch value, when the material becomes transparent for the
pump wave.

\section{Basic equations}

The system of  equations describing three wave interactions (one
dimensional case) in a $\chi^{2}$ - medium for the  slowly varying
envelope and phase approximation can be written in the following
form~\cite{report,Shen:84}:
\begin{align}
&\left(\widehat{k}_{1}\frac{\partial}{\partial z} + \frac{1}{\vartheta_{1}} \frac{\partial }{\partial t}\right)A_{1}=\imath \frac{2\pi \omega_{1}^{2}\mu(\omega_{1})}{c^{2}k_{1}}P^{NL}(\omega_{1})\exp{(-\imath k_{1}z)}\notag\\
&\left(\widehat{k}_{2}\frac{\partial}{\partial z}+\frac{1}{\vartheta_{2}}\frac{\partial }{\partial t}\right)A_{2}=\imath \frac{2\pi \omega_{2}^{2}\mu(\omega_{2})}{c^{2}k_{2}}P^{NL}(\omega_{2})\exp{(-\imath k_{2}z)} \label{main}\\
&\left(\widehat{k}_{3}\frac{\partial}{\partial z}+\frac{1}{\vartheta_{3}}\frac{\partial }{\partial t}\right)A_{3}=\imath \frac{2\pi \omega_{3}^{2} \mu(\omega_{3})}{c^{2}k_{3}}P^{NL}(\omega_{3})\exp{(-\imath k_{3}z)}\notag
\end{align}
where wave numbers $k_{j},~j=1,2$ are defined as follows
$k_{j}^{2}=\left(\omega_{j}/c\right)^{2}\varepsilon(\omega_{j})\mu(\omega_{j})$
and $\widehat{k}_{j}$ is the sign of the square root of
$n_{j}^{2}=\varepsilon(\omega_{j})\mu(\omega_{j})$ and
\begin{align}
&P^{NL}(\omega_{1})=\chi^{2}(\omega_{1};\omega_{3},-\omega_{2})A_{3}A^{*}_{2}\exp{(\imath z (k_{3}-k_{2}))}\notag\\
&P^{NL}(\omega_{2})=\chi^{2}(\omega_{2};\omega_{3},-\omega_{1})A_{3}A^{*}_{1}\exp{(\imath z (k_{3}-k_{1}))}\\
&P^{NL}(\omega_{3})=\chi^{2}(\omega_{3};\omega_{1},\omega_{2})A_{1}A_{2}\exp{(\imath z (k_{1}+k_{2}))}\notag
\end{align}
For the case of second harmonic generation Eqs.~\eqref{main} take
the following form:
\begin{align}
&\left(\widehat{k}_{\omega}\frac{\partial}{\partial z}+
\frac{1}{\vartheta_{\omega}}\frac{\partial }{\partial t}\right)
A_{\omega}=\imath \frac{2\pi \omega^{2}
\chi^{2}(\omega)\mu(\omega)}{c^{2}k_{\omega}}A_{2\omega}
A_{\omega}^{*}\exp{(-\imath \Delta k z)}\notag\\
&\label{SHG}\\
&\left(\widehat{k}_{2\omega}\frac{\partial}{\partial z}+
\frac{1}{\vartheta_{2\omega}}\frac{\partial }{\partial t}\right)
A_{2\omega}=\imath \frac{2\pi (2\omega)^{2}\chi^{2}(2\omega)\mu(2\omega)}{c^{2}k_{2\omega}}A_{\omega}^{2}\exp{(\imath \Delta k z)}\notag
\end{align}
where $\Delta k=2k_{\omega}-k_{2\omega}$, $A_{\omega}$ is the
fundamental wave with frequency $\omega$, and a$A_{2\omega}$ is
the second harmonic generated in the material. We consider the case,
when the refractive index is negative at the fundamental frequency
$\omega$ and is positive at the second-harmonic frequency
$2\omega$. The parameter $\Delta k$ plays an important role for the spatial
distribution of the electromagnetic field along the sample. In the
next section we consider both cases $\Delta k=0$ and $\Delta k \ne
0$.

\section{Case of ideal phase matching  $\Delta k=0$}
\begin{figure}[h]
\noindent\centering{
\includegraphics[width=90 mm]{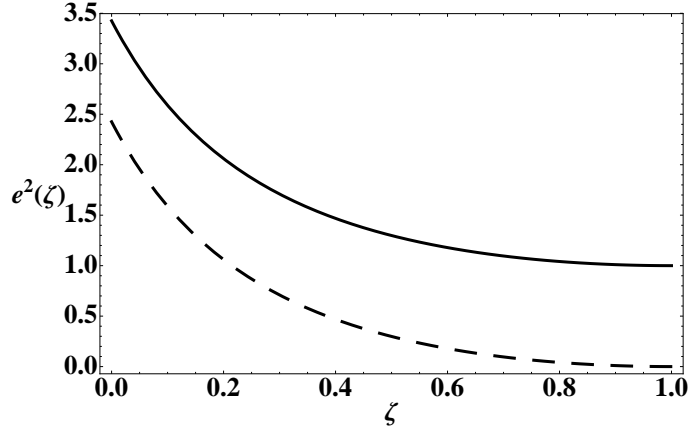}}
\caption{The dependence of the intensity of fundamental wave
$e_{1}^{2}$ (solid curve) and second harmonic $e_{2}^{2}$
(dashed curve) on the distance $\zeta$ with $e_{10}^{2}=3.5$ }
\label{figEvsZ}
\end{figure}

We  consider second harmonic generation for continuos waves in a
$\chi^{2}$ medium under ideal phase matching conditions $\Delta k=0$.
The length of the sample we assume to be $L$. Using the symmetry
properties of the susceptibility tensor $\chi^{2}$ with respect to
permutations of $\omega$ and $2\omega$ frequencies, the mathematical
model  of second harmonic generation can be formulated in the
following way~\cite{Shen:84,Boyd:92}:
\begin{align}
&\frac{d A_{\omega}}{d z}=-\imath \frac{2K\omega^{2} \mu(\omega)}{c^{2}
k_{\omega}}A_{2\omega}A_{\omega}^{*}\label{SHGCONT2}\\
&\frac{d A_{2\omega}}{d z}=\imath \frac{4K\omega^{2} \mu(2\omega)}{c^{2}
k_{2\omega}}A_{\omega}^{2}\label{SHGCONT3},\\
&A_{\omega}(0)=A_{\omega}^{0},~~A_{2\omega}(L)=0,
\end{align}
where $K=2\pi \chi^{2}(2\omega)/c^{2}=\pi \chi^{2}(\omega)/c^{2}$.
Let us represent the complex functions $A_{\omega}$ and $A_{2\omega}$
in terms of amplitudes $e_{1,2}$ and phases $\varphi_{1,2}$
\begin{eqnarray}
A_{\omega}=e_{1}\exp{(\imath \varphi_{1})}  \mspace{18mu}   A_{2\omega}=
e_{2}\exp{(\imath \varphi_{2})} \label{CompFun}.
\end{eqnarray}
Substitution of Eqs.~\eqref{CompFun} into~\eqref{SHGCONT2}
and~\eqref{SHGCONT3} , and separation of real and imaginary parts
lead to the  following system of equations:
\begin{align}
&\frac{d e_{1}}{d z}=\kappa e_{1}e_{2}\sin{(\theta)},\notag\\
&\frac{d e_{2}}{d z}=\kappa e_{1}^{2}\sin{(\theta)},\label{SHGContMain}\\
&\frac{d \theta}{d z}=\kappa\left(\frac{e_{1}^{2}}{e_{2}}+2e_{2}\right)\cos{(\theta)},\notag
\end{align}
with boundary conditions:
\begin{equation}
e_{1}(0)=e_{10},~~e_{2}(L)=0.\label{Boundary:cond}
\end{equation}
Here $\theta$ and $\kappa$ are defined as follows
\begin{equation}
\theta=\varphi_{2}-2\varphi_{1},~~ \kappa=4K\omega^{2} \mu(2 \omega)/c^{2}k_{2 \omega}, \label{def:kappa:and:theta}
\end{equation}
From the first two equations an integral of motion follows:
\begin{equation}
e_{1}^{2}-e_{2}^{2}=m_{1}^{2}=const\label{Manley:Row}
\end{equation}
This integral of motion corresponds to the modified Manley-Row
relation. In  case of second harmonic generation in conventional
materials the Manley-Row relation is equivalent to conservation of
energy ($e_{1}^{2}+e_{2}^{2}=const$). In our case,
relation~\eqref{Manley:Row} corresponds to conservation of total
flux of the energy.  The second integral of motion for the
system~\eqref{SHGContMain} reads as:
\begin{equation}
e_{1}^{2}e_{2}\cos{(\theta)}=m_{2}=const.\label{Second:integral}
\end{equation}
The integral of motion~\eqref{Second:integral} is consistent with
boundary  conditions~\eqref{Boundary:cond} only if
$\cos{(\theta)}=0$. Taking into account that the pump wave energy
decays in $z$, we conclude that the phase difference is equal to
$\theta=3\pi/2$, therefore the system of equations~\eqref{SHGContMain}
can be represented as follows:
\begin{eqnarray}
\frac{d e_{1}}{d z}=-\kappa e_{1}e_{2}, \mspace{18 mu} \frac{d
e_{2}}{d z}=-\kappa e_{1}^{2} \label{SHGContFin}
\end{eqnarray}
The solution of~\eqref{SHGContFin} has the following form
\begin{align}
&e_{1}(\zeta)=m_{1}/\cos{(m_{1}(\textit{l}-\zeta))}\notag\\
&\label{solutions:e1:e2}\\
&e_{2}(\zeta)=m_{1}\tan{(m_{1}(\textit{l}-\zeta))}\notag
\end{align}
here $\zeta=\kappa z$ and $\textit{l}=\kappa L$.
The solutions~\eqref{solutions:e1:e2} unknown parameter $m_1$,
is the value of the fundamental field at the end of the sample.
This parameter can be found from the boundary
condition~\eqref{Boundary:cond}. Taking into account the Manley-Row
relation~\eqref{Manley:Row}, it leads to the transcendental
equation for $m_{1}$:
\begin{equation}
e_{10}=m_{1}/\cos(m_{1}l).
\label{eq:trans:m1}
\end{equation}
This equation can be solved numerically. The solution
of~\eqref{eq:trans:m1} together with~\eqref{solutions:e1:e2} determines
the field distribution along the sample.  The dependence of
intensities $e_{1}^{2}$ and $e_{2}^{2}$ on $\zeta$ is represented in
Fig.\ref{figEvsZ} were the intensity boundary value
$e_{1}^{2}(0)$ is chosen to be $e_{10}^{2}=3.5$, here $\textit{l}=1$
and $m_{1}=1$.

\begin{figure}[h]
\noindent\centering{
\includegraphics[width=90 mm]{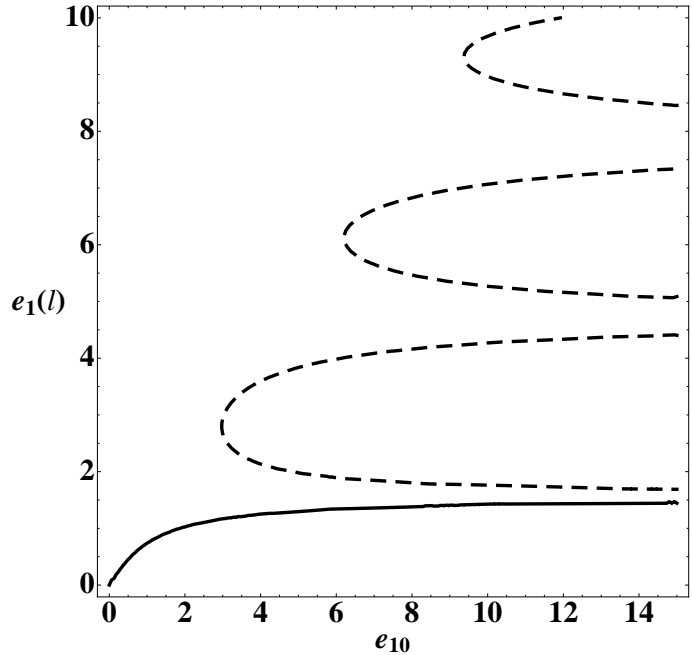}
} \caption{The dependence of the intensity of output fundamental
wave $e_{1}(l)$ on the $e_{10}$} \label{figMvsE}
\end{figure}

\begin{figure}[h]
\noindent\centering{
\includegraphics[width=90 mm]{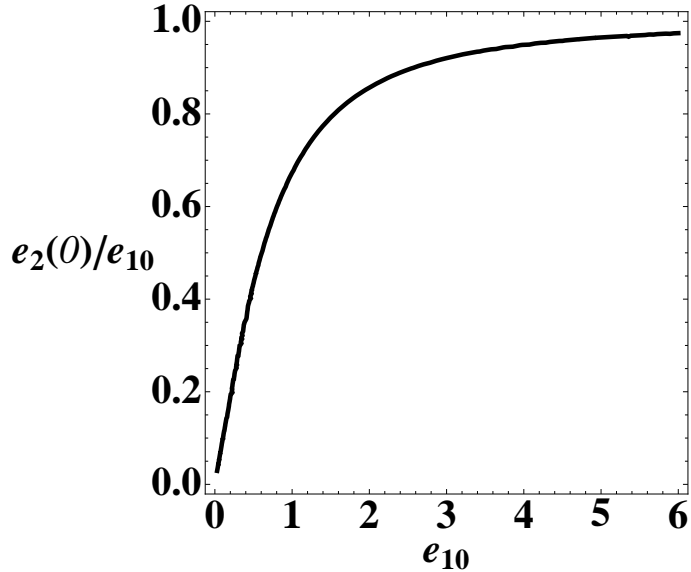} }
\caption{The dependence of conversion efficiency of the pump to
second harmonic fields. } \label{fig:second:harm:convers}
\end{figure}

The solution of transcendental  equation ~\eqref{eq:trans:m1} for
$l=1$ is shown in Fig.\ref{figMvsE}. This plot illustrates the
dependence of the output field intensity  $e_{1}(l)=m_{1}$, as a function
of $e_{10}$ (the amplitude of the fundamental field pumped into the
medium). As shown in  Fig.\ref{figMvsE}, the formal solution of
equation~\eqref{eq:trans:m1} has multiple branches. However,
only the lower branch presented by a solid curve has physical meaning.
Upper brunches represented by dashed curves are originated from
periodicity of the $\cos$ function in~\eqref{eq:trans:m1}. Both $e_{1}(\zeta)$
and $e_{1}(\zeta)$ corresponding to these branches have
singularities on the interval $0\le \zeta \le l$ which is
inconsistent with conservation of energy. Note that the lower ``physical"
branch shows saturation of output power of the electric field at
the fundamental frequency $e_{1}(l)$ with increase of
input power $e_{1}(0)$. This indicates that with the increase of
input power $e_{1}(0)$  above $2$, all excessive energy of pump
signal converts to energy of the second harmonic signal (see
Fig.~\ref{fig:second:harm:convers}).
\section{Second harmonic generation in the
presence of a phase   mismatch $\Delta k \ne 0$}
\begin{figure}[h]
\noindent\centering{
\includegraphics[width=90 mm]{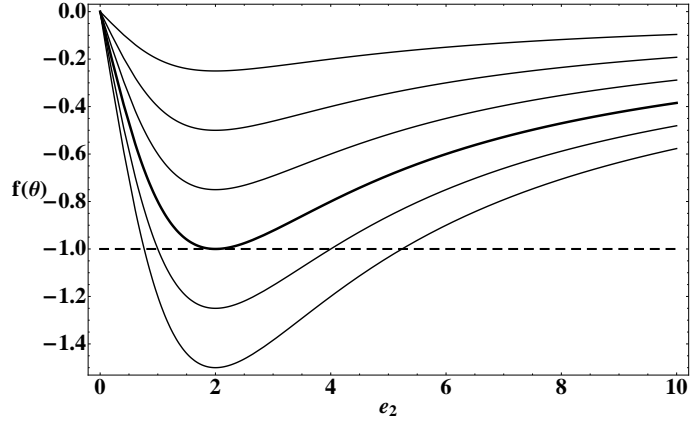}
} \caption{The dependence of the $\cos{(\theta)}$ on $e_{2}$ with
different values of $\Delta=k \times m_1$, here $k=1, 2,\ldots, 6$
and $m_{1}=1.2$. The bold curve corresponds to a critical value of
 $\Delta=\Delta_{cr}$} \label{figThetavsE2}
\end{figure}

Let us consider the impact of phase mismatch $\Delta k$ (Eq.
\eqref{SHG}) on second harmonic generation. The system of equations
describing the spatial distribution of field amplitudes $e_{1,2}(z)$ and
phase difference $\theta(z)$ in the presence of phase mismatch reads:
\begin{align}
&\frac{de_{1}}{d z}=\kappa e_{1}e_{2}\sin{(\theta)},\notag\\
&\frac{de_{2}}{d z}=\kappa e_{1}^{2}\sin{(\theta)},\label{SHGdelat1}\\
&\frac{d \theta}{d z}=\kappa\left(\frac{e_{1}^{2}}{e_{2}}+2e_{2}\right)\cos{(\theta)}-\Delta k.\notag
\end{align}
Here $\theta=\varphi_{2}-2\varphi_{1}-\Delta kz $ and $\kappa$ is
defined in~\eqref{def:kappa:and:theta}. By introducing variables
$\zeta=\kappa z$ and $\textit{l}=\kappa L$  Eqs.~\eqref{SHGdelat1}
can be represented in the following form:
\begin{align}
&\frac{de_{1}}{d \zeta}= e_{1}e_{2}\sin{(\theta)},\notag\\
&\frac{de_{2}}{d \zeta}= e_{1}^{2}\sin{(\theta)},\label{SHGdelat2}\\
&\frac{d \theta}{d \zeta}=
\left(\frac{e_{1}^{2}}{e_{2}}+2e_{2}\right)\cos{(\theta)}-\Delta,\notag
\end{align}
here $\Delta= \Delta k/\kappa$.  The Manley-Row relation in this
case remains unchanged:
\begin{equation}
e_{1}^{2}-e_{2}^{2}=m_{1}^{2}=const,~~m_{1}= e_{1}(l).\nonumber
\end{equation}
and a second integral of motion in presence of phase mismatch reads:
\begin{equation}
e_{2}e_{1}^{2}\cos{(\theta)}+\frac{e_{2}^2\Delta}{2}=m_{2}=const
\end{equation}
Taking into account boundary condition   $e_{2}(l)=0$, we conclude
that $m_{2}=0$ and therefore
\begin{equation}
\cos{(\theta)}=-\frac{\Delta }{2}\frac{e_{2}}{(m_{1}^2+e_{2}^{2})}
\label{cos}
\end{equation}
The function~\eqref{cos} has an extremum at  $e_{2}^{2}=m_{1}^{2}$ and
$\cos{(\theta)}$ at this value of $e_{2}$ gives
$\cos{(\theta)}=-\Delta/4 m_{1}$. Since
 $|\cos{(\theta)}|\le 1$,  then there is the critical value of mismatch
 $ |\Delta_{cr}|=4  m_{1}$  such that $\max|\cos \theta| =1$.
Notice that~\eqref{cos} is  defined for
 arbitrary values of $e_2$ if $|\Delta|\le 4  m_{1}$. If $|\Delta|\ge 4
 m_{1}$ then there is a forbidden  gap for values of $e_2$:

\begin{equation}
\frac{1}{4}\left(|\Delta|-\sqrt{\Delta^{2} \label{left:e2}
-\Delta_{cr}^{2}}\right) < e_2 <
\frac{1}{4}\left(|\Delta|+\sqrt{\Delta^{2} -\Delta_{cr}^{2}}\right).
\end{equation}
In this case  $|\cos \theta | \le 1$ if
\begin{align}
&e_2 \ge \frac{1}{4}\left(|\Delta|+\sqrt{\Delta^{2} -\Delta_{cr}^{2}}\right) \label{right:e2}\\
&0\le e_2 \le \frac{1}{4}\left(|\Delta|-\sqrt{\Delta^{2}
\label{left:e2} -\Delta_{cr}^{2}}\right).
\end{align}
Since the value of $e_2$ on the right side of the sample is ser to be
zero ($e_{2}(l)=0$), the branch of $e_2$ values~\eqref{right:e2} is
not accessible. Values of $e_2$ in this case remain within
the branch~\eqref{left:e2}. In this case the conversion efficiency of
the pump wave to second harmonic is limited by the value $4
e_{10}/(|\Delta|-\sqrt{\Delta^{2} -16m_{1}^{2}})$. The dependence of
$f(\theta)= -\Delta e_{2}/2(m_{1}^2+e_{2}^{2})$ on
$e_{2}$ for different values of mismatch is shown in
Fig.~\ref{figThetavsE2}. The bold curve on this figure corresponds to a
critical value of the mismatch. The forbidden  gap for $e_{2}$ can be seen
for two lowest curves, when curves are below $-1$.
The presence of a forbidden gap for $e_2$ suggests the existence of two types
of solutions for $e_2$. The first type corresponds to mismatch values
$|\Delta |\le 4m_1$ and in this case $e_2$ is not bounded from above.
This means  that the conversion rate of fundamental harmonic to second
harmonic, in principle, can be high (close to $1$ - ideal
conversion, similar to  Fig.~\ref{fig:second:harm:convers} in the
previous section).  The second type of solutions correspond to mismatch
values $|\Delta |\ge 4m_1$; in this case the amplitude $e_2$ is bounded
from above $0\le 4 e_{2}\le |\Delta|-\sqrt{\Delta^{2}
-16m_{1}^{2}}$. This means the is a limitation  of the output intensity of
second harmonic field at the growing input intensity of fundamental
harmonic.

For further considerations it is more convenient to deal with field
intensities rather then with amplitudes. Using
expression~\eqref{cos} for $\cos{(\theta)}$, the second equation of
~(\ref{SHGdelat2}) can be represented as an equation for the intensity
$P_{2}=e_{2}^{2}$:
\begin{align}
&\frac{d P_{2}}{d \zeta}=\left\lbrace F(P_{2})\right\rbrace^{1/2},
\label{P2:delta}
\end{align}
where $F(P_{2})$ is a quartic polynomial
\begin{align}
&F(P_{2})=4 P_{2}^{3}+\left(8 m_{1}^{2}-\Delta^{2}\right) P_{2}^{2} + 4 m_{1}^{4} P_{2}.\notag
\end{align}
with the following roots:
\begin{align}
&P_{2c}=\frac{1}{8}\left(\Delta^{2}-8m_{1}^{2}+\Delta
\sqrt{\Delta^{2}-16 m_{1}^{2}}\right)\nonumber\\
&P_{2b}=\frac{1}{8}\left(\Delta^{2}-8m_{1}^{2}-\Delta
\sqrt{\Delta^{2}-16 m_{1}^{2}}\right)\label{roots:main}\\
&P_{2a}=0. \nonumber
\end{align}
Notice that these roots~\eqref{roots:main} define the forbidden gap
$[\sqrt{P_{2b}},\sqrt{P_{2c}}]$ for values of $e_2$
(see Fig.~\ref{figThetavsE2}
and equations~\eqref{right:e2},~\eqref{left:e2}).

Based on this qualitative analysis, we conclude that there are three
regimes of second harmonic generation controlled by the absolute
value of the phase mismatch. In the following subsection we will
analyse solutions describing spatial field distribution inside
the sample.

\subsection{Three regimes of second harmonic generation }
The absolute value of the phase mismatch determine three different
regimes of second harmonic generation: $|\Delta |< \Delta _{cr}$,
$|\Delta |= \Delta _{cr}$ and $| \Delta |> \Delta _{cr}$. First we
consider the case of subcritical mismatch: $| \Delta |< \Delta _{cr}$.
\subsubsection{Subcritical mismatch}
In the case where $|\Delta| <\Delta_{cr}$, roots~\eqref{roots:main} are
complex-valued and the solution of~\eqref{solution:gen} can be
expressed in terms of Weierstrass  function $\wp$~\cite{Whitt:88}.
By expanding $F\left(P_{2}\right)$ into Taylor series and
introducing a new variable:
\begin{align}
&s=\frac{F'(P_{2a})}{4(P_{2}-P_{2a})}+\frac{1}{24}F''(P_{2a}).\notag
\end{align}
Here derivatives of the polynomial are taken with respect to
$P_{2}$,  then the solution of equation~(\ref{P2:delta}) can be represented
in an implicit form:
\begin{align}
& \zeta - \textit{l} = \int_{s}^{\infty} \frac{d s}{\left\lbrace 4 s^{3} -
g_{2} s - g_{3}\right\rbrace^{1/2} }, \label{solution:gen}
\end{align}
where $g_{2}$ and $g_{3}$ are invariants of Weierstrass function.
\begin{align}
&g_{2}=\frac{1}{12} \left(8 m_{1}^{2}-\Delta ^2\right)^2-4
m_{1}^{4}, \mspace{18 mu} g_{3}=\frac{1}{3} m_{1}^{4} \left(8
m_{1}^{2}-\Delta ^2\right)-\frac{1}{216} \left(8 m_{1}^{2}-\Delta
^2\right)^3.\notag
\end{align}
\begin{figure}[t]
 \noindent\centering{
\includegraphics[width=90 mm]{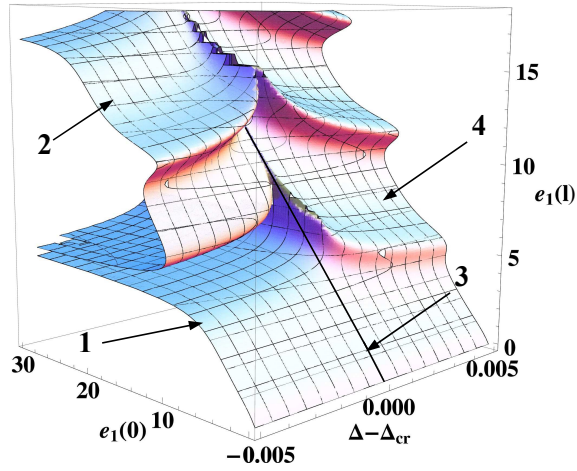}
} \caption{The dependence of the output field  amplitude $e_{1}(l)$
at the fundamental frequency on $e_{10}$ near the critical value of phase
mismatch $\Delta_{cr}$} \label{fig:contour:3D}
\end{figure}
Finally, the amplitudes of second and fundamental harmonics have
the following form:
\begin{align}
&e_{1}(\zeta)=\sqrt{m_{1}^{2}+e_{2}^{2}(\zeta)},\label{e1:Delta:solution}\\
&e_{2}(\zeta)=\frac{m_{1}^{2}}{\sqrt{
\left(\wp{(\textit{l}-\zeta;g_2,g_3)}-\left(8
m_{1}^{2}-\Delta^{2}\right)/12\right)}}\label{e2:Delta:solution}
\end{align}

The parameters $g_{2}$ and $g_{3}$ are functions of $\Delta$ and
$m_{1}$.  To determine solutions of Eqs.~(\ref{SHGdelat2}) we need to
solve for the unknown value of the output pump wave  $m_1$. The value of
$m_1$ can be found taking into account the output boundary condition
$m_{1} = e_1(l)$ and the Manley-Row relation~(\ref{Manley:Row}), which
lead to the following transcendental equation for $m_1$:
\begin{align}
&e_{10}^{2}=m_{1}^{2}+\frac{m_{1}^{4}}{
\wp{(\textit{l};g_2,g_3)}-\left(8 m_{1}^{2}-\Delta^{2}\right)/12}
\label{eq:trans:gen}
\end{align}

\begin{figure}[t]
\noindent\centering{
\includegraphics[width=90 mm]{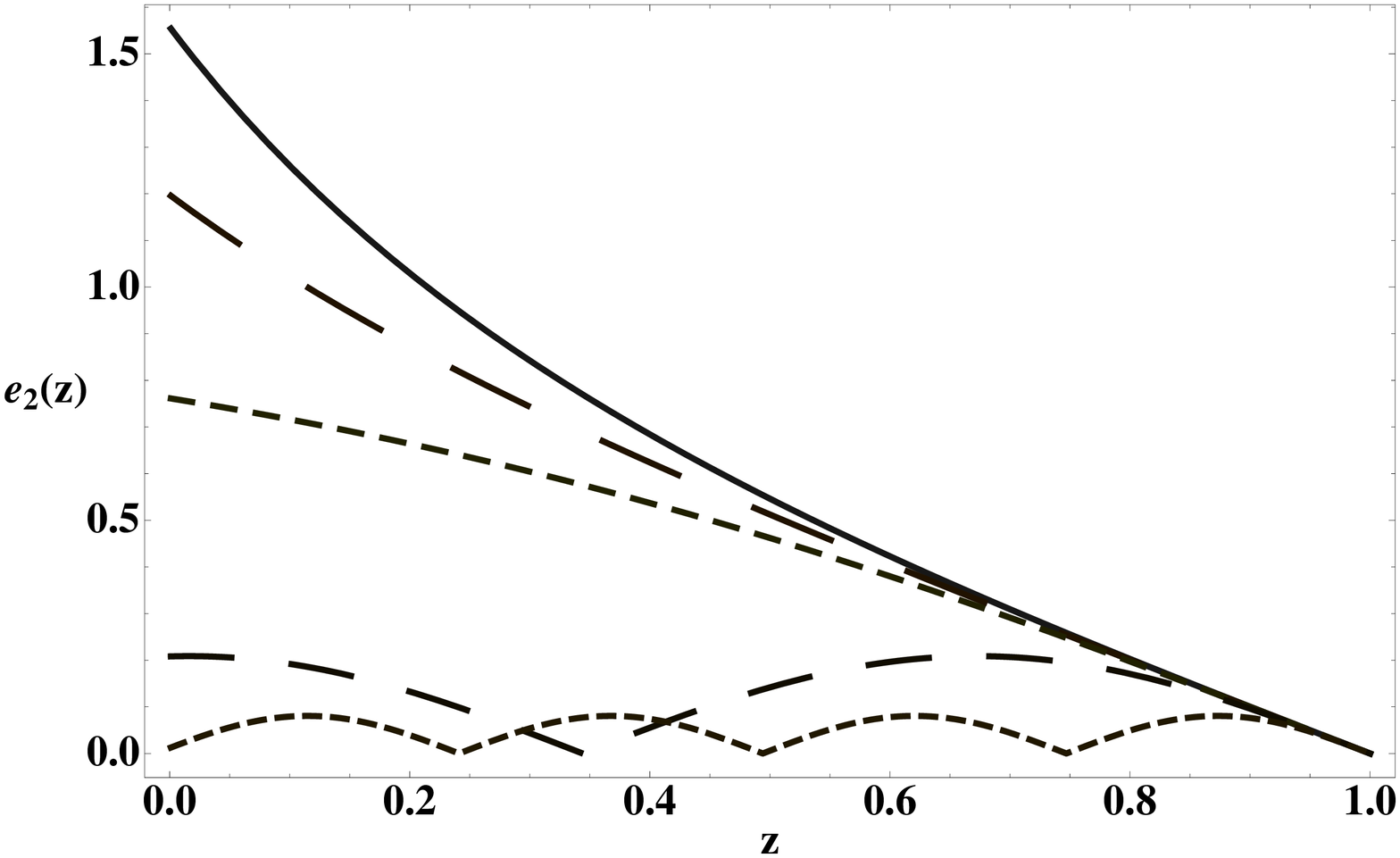}
} \caption{The dependence of second harmonic's amplitude
$e_{2}(\zeta)$ on the $\zeta$ with different values of phase
mismatch(the solid curve: $\Delta=0$, large dashed curve $\Delta=2
m_{1}$, small dashed curve: $\Delta_{cr}=4 m_{1}$, intermediate
dashed curve: $\Delta=10 m_{1}$ , doted curve: $\Delta=25 m_{1}$). }
\label{figE2vsD}
\end{figure}

To determine the unknown parameter $m_{1}=e_{1}(l)$,
Eq.\eqref{eq:trans:gen} needs to be solved numerically. The analysis
of the $e_{1}(l)$ dependence on $e_{10}$ and $\Delta$ shows that with
increasing phase mismatch from $0$ to $\Delta_{cr}$, all branches
(physical and nonphysical) shift upwards and nonphysical branches
change their shapes. Fig.~\ref{fig:contour:3D} shows the dependence
of the output field amplitude $e_{1}(l)$ at fundamental frequency on
$e_{10}$ near the critical value of phase mismatch $\Delta_{cr}$. Sheet,
labeled as ``1", corresponds to the physical branch, while sheet labeled
as ``2" represents the first nonphysical branch. Other nonphysical
sheets are located above nonphysical sheet ``2" shown on
Fig.~\ref{fig:contour:3D}. The spatial distribution of $e_{1}(\zeta)$
and $e_{2}(\zeta)$ can be found by substitution of the solution of
the Eq.~\eqref{eq:trans:gen} ($m_1$) in
Eqs.~\eqref{e1:Delta:solution} and  \eqref{e2:Delta:solution}. We
found that all solutions $e_{1}(\zeta)$ and $e_{2}(\zeta)$ are
monotonically decreasing in $\zeta$. An example of $e_{2}(\zeta)$ at
$\Delta = \Delta_{cr}/2$ is shown in  Fig.~\ref{figE2vsD}).

The conversion efficiency  $\alpha=e_{2}(l)/e_{10}$ as a function of
the input amplitude $e_{10}$ is presented in
Fig.~\ref{fig:conver:eff:delta}. As one can observe,
$\alpha$ is approaching its
asymptotic value $\alpha = 1$ in slower fashion for larger values of
$|\Delta|$.
\subsubsection{Critical mismatch}
\begin{figure}[t]
 \noindent\centering{
\includegraphics[width=90 mm]{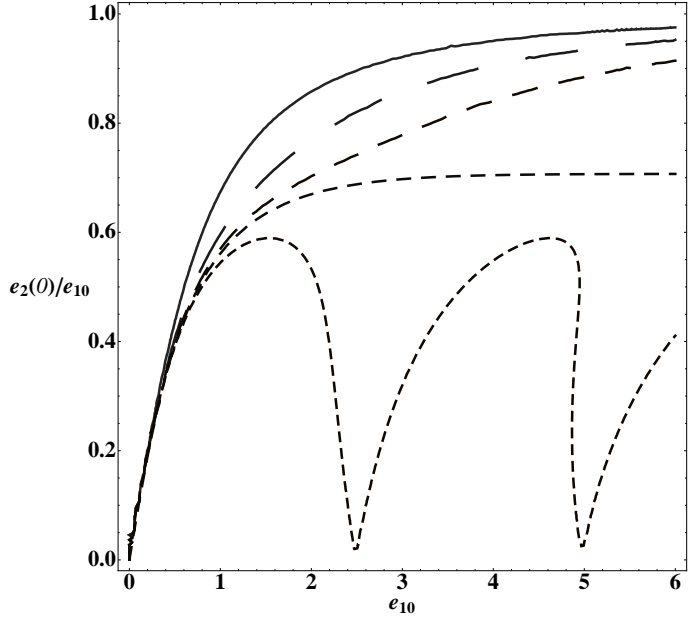}
} \caption{Dependence of conversion efficiency
$\alpha=e_{2}(0)/e_{10}$ on input field amplitude $e_{10}$ with
different values of phase mismatch $\Delta$: Solid curve
$\Delta=0$; large dashed curve $\Delta=3.5 m_{1}$; dashed curve
$\Delta=3.9 m_{1}$; small dashed curve $\Delta_{cr}=4 m_{1}$; doted
oscillation curve $\Delta=4.5 m_{1}> \Delta_{cr}$}
\label{fig:conver:eff:delta}
\end{figure}

When  the value of phase mismatch is critical $|\Delta|=\Delta_{cr}$,
the roots $P_{2c}=P_{2b}=\Delta_{crit}^{2}/16=m_{1}^{2}$ and
the discriminant of the Weierstrass function is zero. In this case, the
 function
$\wp(\textit{l}-\zeta;g_{2},g_{3})$ can be represented in terms of
hyperbolic functions. Thus the Eqs.~\eqref{e1:Delta:solution}
and~\eqref{e2:Delta:solution} take the form:
\begin{align}
&e_{1}(\zeta)=\sqrt{m_{1}^{2}+e_{2}^{2}(\zeta)}\\
&e_{2}(\zeta)=m_{1}\tanh{(m_{1}(\textit{l}-\zeta))}
\end{align}
and the transcendental equation for $m_{1}$ reads as
\begin{align}
&e_{10}^{2}=m_{1}^{2}\left(1+\tanh^{2}{(m_{1}\textit{l})}\right)
\label{eq:trans:crit}
\end{align}
The
numerical solution of Eq.~\eqref{eq:trans:crit} is shown in
Fig.\ref{fig:contour:3D} (line ``3"). Observe that at large values of
$e_{10}$, the solution of Eq.~\eqref{eq:trans:crit} is proportional to
$e_{10}$ ($m_{1} \approx e_{10}$). Therefore, at large values of
$e_{10}$ the conversion efficiency  $\alpha \rightarrow \tanh(l e_{10})$
is always less then one while in the subcritical regime $\alpha
\rightarrow 1$ (see Fig.~\ref{fig:conver:eff:delta}).
\subsubsection{Overcritical mismatch}
\begin{figure}[t]
 \noindent\centering{
\includegraphics[width=90 mm]{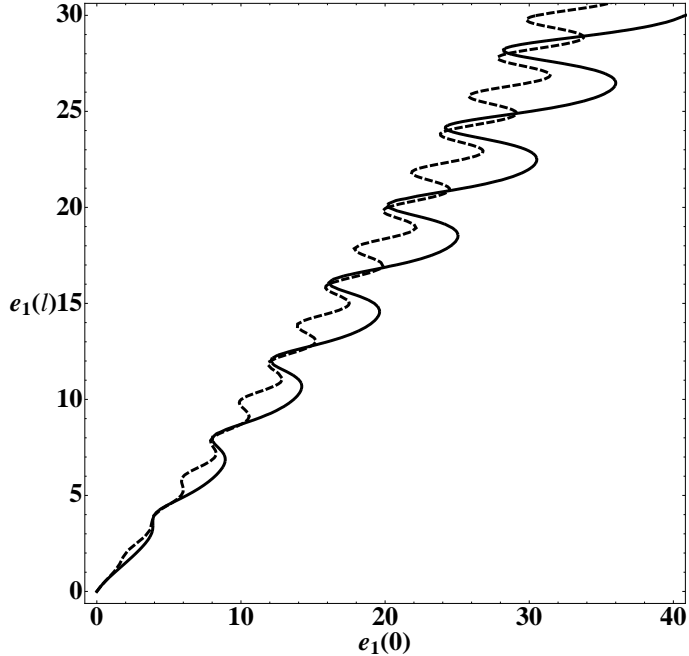}
} \caption{The dependence of the output field  amplitude $e_{1}(l)$
at fundamental frequency on $e_{10}$. Solid curve $\Delta=4.1
m_{1}$, dashed curve $\Delta=4.5 m_{1}$}
\label{fig:bistab}
\end{figure}

At large mismatch values,  when  $|\Delta | >\Delta_{cr}$, roots
of~\eqref{roots:main} are real.  In this case it is convenient to
represent $\wp(\textit{l}-\zeta;g_{2},g_{3})$ in terms of Jacobi
elliptic $sn$
 function~\cite{Shabat:51}.
Eqs~\eqref{e1:Delta:solution} and ~\eqref{e2:Delta:solution} can be
represented as
\begin{align}
&e_{1}(\zeta)=\sqrt{m_{1}^{2}+e_{2}^{2}(\zeta)}\label{e1:above:solution}\\
&e_{2}(\zeta)=\sqrt{P_{2b}} ~\mathrm{sn}\left[\sqrt{P_{2c}}(\textit{l}-\zeta),\gamma\right],
\label{e2:above:solution}
\end{align}
here $\gamma=\sqrt{P_{2b}/P_{2c}}$, and the equation for $m_{1}$ takes
the following form:
\begin{align}
&e_{10}^{2}=m_{1}^{2}+P_{2b} ~\mathrm{sn}^{2}\left[\sqrt{P_{2c}}\textit{l},\gamma\right]
\label{eq:trans:overcrit}
\end{align}
The sheet corresponding to solutions of~\eqref{eq:trans:overcrit} is
labeled in Fig.~\ref{fig:contour:3D}  as ``4". In contrast to the
subcritical regime, all solutions in this case are represented by a
single sheet. This  sheet has folds. Hence the intersection of this
sheet with plane corresponding to $\Delta = const$ gives multivalued
dependance of  $e_{1}(l)$ on $e_{10}$. This dependance for two
different values of $\Delta$ is shown in Fig.~\ref{fig:bistab}.

\begin{figure}[t]
\noindent\centering{
\includegraphics[width=90 mm]{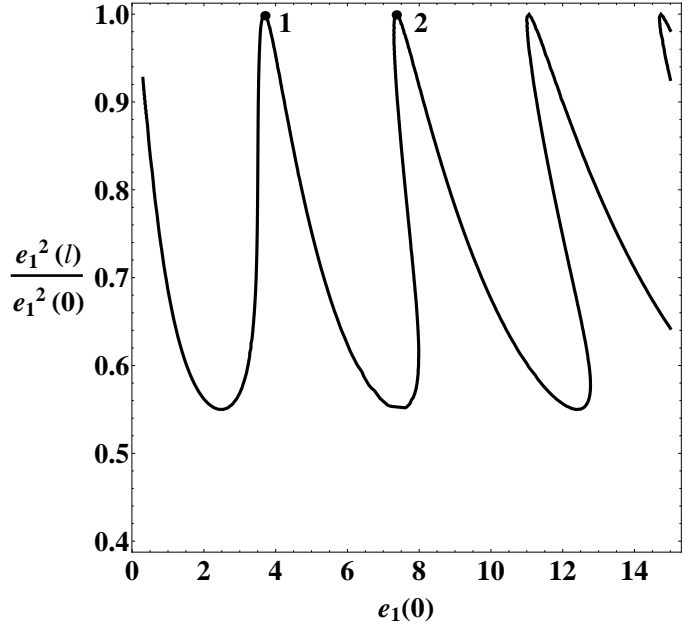}
} \caption{Dependence of the transmission coefficient
$\Im$ on pumped field amplitude $e_{10}$. $\Delta=4.2 m_{1}$ }
\label{fig:trans:coeff}
\end{figure}

\begin{figure}[t]
\noindent\centering{
\includegraphics[width=90 mm]{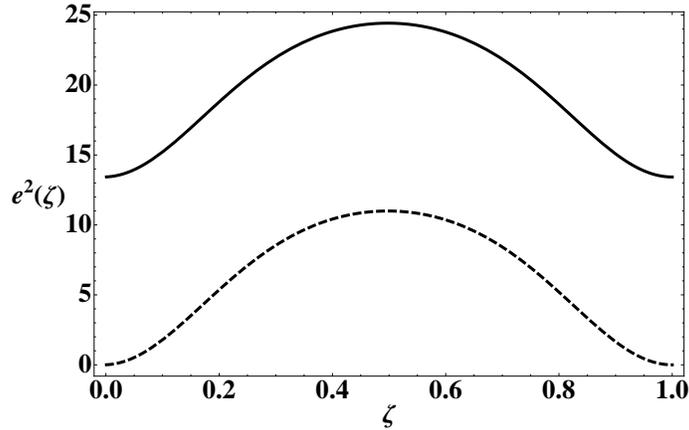}
} \caption{Spatial distribution of the intensities  $e_{1}^{2}(\zeta)$
(solid curve) and $e_{2}^{2}(\zeta)$ (dashed curve) inside the slab}
\label{fig:point:1}
\end{figure}

In the supercritical regime the second harmonic field experiences spatial
periodic oscillations  with period $4 \textbf{K}(\gamma)$ (see
Fig.~\ref{figE2vsD}). The distance between neighboring zeros
$\tilde{\zeta}$ of the amplitude of second harmonic is determined by the
following formula:
\begin{align}
\tilde{\zeta}=\frac{2\textbf{K}(\gamma)}{\sqrt{P_{2c}}}
\end{align}
If the slab length satisfies the condition $l=n \times
\tilde{\zeta}$ ($n=1,2,3\ldots$), then the amplitude  of the second harmonic
wave is zero at the both ends of the slab (zero conversion
efficiency). Therefore  such slab  is transparent for a pump wave.
A plot of the transmission coefficient
$\Im=e_{1}(\textit{l})^{2}/e_{10}^{2}$ as function of  $e_{10}$ is
shown in Fig.\ref{fig:trans:coeff}. The  transmission coefficient is
equal to $1$ at the points labeled as ``1", ``2", $\ldots$
(transmission resonances).  The spatial distribution of the
fundamental and second harmonic fields corresponding to the transmission
resonance at the point ``1" (see Fig.\ref{fig:trans:coeff}) is shown
in Fig.\ref{fig:point:1}.

\section{Conclusion}

We considered second harmonic generation in negative index
materials. Specifics of this process is in  the negative value of
refractive index for the pump wave and the positive value for second
harmonic. This led to important features which are different  from
the case of  second harmonic generation in conventional dielectrics.
The main difference is in the existence of nonzero critical values of
the phase mismatch. If the absolute value of phase mismatch is below
critical, then  the field intensities are monotonically decaying
along the sample. When the absolute value of phase mismatch exceeds
a critical value, monotonic decay of intensities transforms to a
spatial periodic oscillations. Note, that in the conventional case
the critical value of phase mismatch is zero.

Another important feature is the dependance of conversion efficiency
on the amplitude of the incident pump wave. When the absolute value of
phase mismatch is below critical value, then the conversion efficiency
asymptotically approaches 100\% at large values of the incident pump
wave amplitude. It should be stressed that in this case the
asymptotic value of conversion efficiency does not depend on the phase
mismatch value. The phase mismatch affects only the rate of approaching
of conversion efficiency to its asymptotic value. When the phase
mismatch is exactly equal to critical value, then the asymptotic
value of conversion efficiency experiences a jump to a value which
is less than 100\%. When the absolute value of phase mismatch is above
critical value, the conversion efficiency becomes an oscillatory function
of the incident pump wave amplitude.

Finally, we found that the dependance of output amplitude of the pump
wave on its input amplitude is single valued if the absolute value of
phase mismatch is below critical and becomes multi-valued in the
opposite case.

\section*{Acknowledgments}
We would like to thank  A. K. Popov and V. M. Shalaev  for valuable
discussions and A. Aceves for  help during preparation of this
paper.  A.I.M and Zh.K. appreciate  support and hospitality of the
University of Arizona Department of Mathematics during the
preparation on this manuscript. This work was partially supported by
NSF (grant DMS-0509589), ARO-MURI award 50342-PH-MUR and State of
Arizona (Proposition 301), RFBR (grant No. 09-02-00701-a) and the
Federal Goal-Oriented Program ``Scientific and
Scientific-Educational Personnel of Innovational Russia".

\end{document}